\documentclass[aps,prx,superscriptaddress,showpacs,reprint]{revtex4-1}
\usepackage{amsmath}
\usepackage{amssymb}
\usepackage{setspace}
\usepackage{graphicx}
\usepackage{subfigure}
\usepackage{braket}
\usepackage{mathrsfs}
\usepackage[colorlinks = true,linkcolor = red,urlcolor  = blue,citecolor = blue,anchorcolor = blue]{hyperref}
\begin{document}

\title{Role of dissipation in realistic Majorana nanowires}
\author{Chun-Xiao Liu}
\affiliation{Condensed Matter Theory Center and Joint Quantum Institute and Station Q Maryland,
Department of Physics, University of Maryland, College Park, Maryland 20742-4111, USA}

\author{Jay D. Sau}
\affiliation{Condensed Matter Theory Center and Joint Quantum Institute and Station Q Maryland,
Department of Physics, University of Maryland, College Park, Maryland 20742-4111, USA}

\author{S. Das Sarma}
\affiliation{Condensed Matter Theory Center and Joint Quantum Institute and Station Q Maryland,
Department of Physics, University of Maryland, College Park, Maryland 20742-4111, USA}

\date{\today}

\begin{abstract}
We carry out a realistic simulation of Majorana nanowires in order to understand the latest high-quality experimental data [H. Zhang \textit{et al}., \href{https://arxiv.org/abs/1603.04069}{arXiv:1603.04069} (2016)]  and, in the process, develop a comprehensive picture for what physical mechanisms may be operational in realistic nanowires leading to discrepancies between minimal theory and experimental observations (e.g., weakness and broadening of the zero-bias peak and breaking of particle-hole symmetry). Our focus is on understanding specific intriguing features in the data, and our goal is to establish matters of principle controlling the physics of the best possible nanowires available in current experiments. We identify  dissipation, finite temperature, multi-sub-band effects, and the finite tunnel barrier as the four most important physical mechanisms controlling the zero-bias conductance peak. Our theoretical results including these realistic effects agree well with the best available experimental data in ballistic nanowires.
\end{abstract}

\maketitle

\section{Introduction}
Semiconductor spin-orbit-coupled nanowires in the presence of proximity-induced superconductivity have been experimentally demonstrated~\cite{Zhang2016Ballistic, Mourik2012Signatures, Das2012Zero, Deng2012Anomalous, Churchill2013Superconductor, Finck2013Anomalous, Albrecht2016Exponential, Chen2016Experimental} to possibly carry localized non-Abelian Majorana zero modes (MZMs) provided suitable conditions involving applied magnetic field, chemical potential, and induced superconducting (SC) gap (and spin-orbit coupling) are satisfied to drive the system into a topological superconducting phase~\cite{Sarma2015Majorana, Stanescu2013Majorana}. In particular, theory predicts~\cite{Sau2010Generic, Lutchyn2010Majorana, Oreg2010Helical, Sau2010Non} that the system would manifest a zero-bias conductance peak (ZBCP) in the topological phase with a quantized zero-temperature differential conductance value of $2e^2/h$ in tunneling transport measurements~\cite{Sengupta2001Midgap}.  Such tunneling experiments by many different groups indeed have shown such a ZBCP under appropriate conditions although the actual conductance values are always substantially (by factors of 5--50) smaller than the expected MZM quantized value.  In addition, the ZBCP is invariably broad essentially encompassing almost the whole spectral gap instead of being sharply localized at zero energy.  These lingering discrepancies between precise theoretical predictions and actual experimental observations (which have now persisted for five years) raise some possible questions on the possible MZM interpretation of the data, and other (more mundane) interpretations have also been put forward in the literature~\cite{Liu2012Zero, Bagrets2012Class, Pikulin2012Zero, Lee2014Spin}.

A recent tunneling experiment by Zhang \emph{et al.}~\cite{Zhang2016Ballistic} in ballistic InSb nanowires in proximity to superconducting NbTiN provides by far the best measured ZBCP in the literature, with the measured ZBCP values reaching almost $0.5e^2/h$ above the background conductance.  In addition, the measured tunneling conductance in Ref.~\cite{Zhang2016Ballistic} shows remarkable qualitative agreement with the theoretical predictions in terms of magnetic field and gate voltage dependence, providing perhaps the strongest phenomenological evidence for the predicted existence of MZMs in nanowires. However, there are still some issues in the data~\cite{Zhang2016Ballistic} which appear to be incompatible with theoretical expectations.  First, the ZBCP is still a factor of 5 smaller than the quantized MZM value in spite of the quoted experimental temperature being very low ($\sim$ 50 mK).  Second, the ZBCP is broad covering essentially all of the topological gap instead of being sharply localized at zero bias.  Third, the measured tunneling conductance manifestly breaks particle-hole (p-h) symmetry, which is considered to be an exact symmetry in superconductors.  Fourth, the data do not reflect the expected``Majorana oscillations''~\cite{Cheng2009Splitting, Cheng2010Tunneling, Sarma2012Splitting, Rainis2013Towards, Prada2012Transport} as a function of magnetic field arising from the overlap of the two MZMs localized at the two ends of the nanowire. In addition, the finite-field topological gap is soft precisely where the ZBCP shows up. It is, therefore, unclear whether the measured tunneling conductance in Ref.~\cite{Zhang2016Ballistic} could be taken as unequivocal evidence in support of the existence of non-Abelian MZMs in nanowires.

In the current paper we carry out a realistic simulation of Majorana nanowires in order to understand the data of Ref.~\cite{Zhang2016Ballistic} and, in the process, develop a comprehensive picture for what physical mechanisms in realistic nanowires may lead to discrepancies between minimal theory and experimental observations (e.g., the breaking of p-h symmetry).  There have been earlier works~\cite{Sarma2012Splitting, Rainis2013Towards, Prada2012Transport, Lutchyn2011Search, Stanescu2011Majorana, Wimmer2011Quantum, Akhmerov2011Quantized, Fulga2011Scattering, Lin2012Zero, Stanescu2013Dimensional, Stanescu2012To, Vuik2016Effects} simulating various realistic aspects of Majorana nanowires, but our work has little overlap with them since our focus is on understanding specific intriguing features in the data of Ref.~\cite{Zhang2016Ballistic}, and our goal is to establish matters of principle controlling the physics of the best possible nanowires available in current experiments. Our reason for focusing on Ref.~\cite{Zhang2016Ballistic} is not only the high quality of its data with the large ZBCP and hard zero-field proximity gap, but also the fact that the ballistic nanowires used in Ref.~\cite{Zhang2016Ballistic} are relatively disorder free, thus eliminating the need to consider extrinsic disorder effects~\cite{Adagideli2014Effects, Brouwer2011Topological, Brouwer2011Probability, Sau2012Experimental, Cole2015Effects, Hui2015Bulk, Takei2013Soft, Sau2013Density, Lutchyn2012Momentum, Lobos2012Interplay}.

\begin{figure*}
\subfigure[\label{fig:optimal}]{
\includegraphics[width=7.5cm, height=11cm]{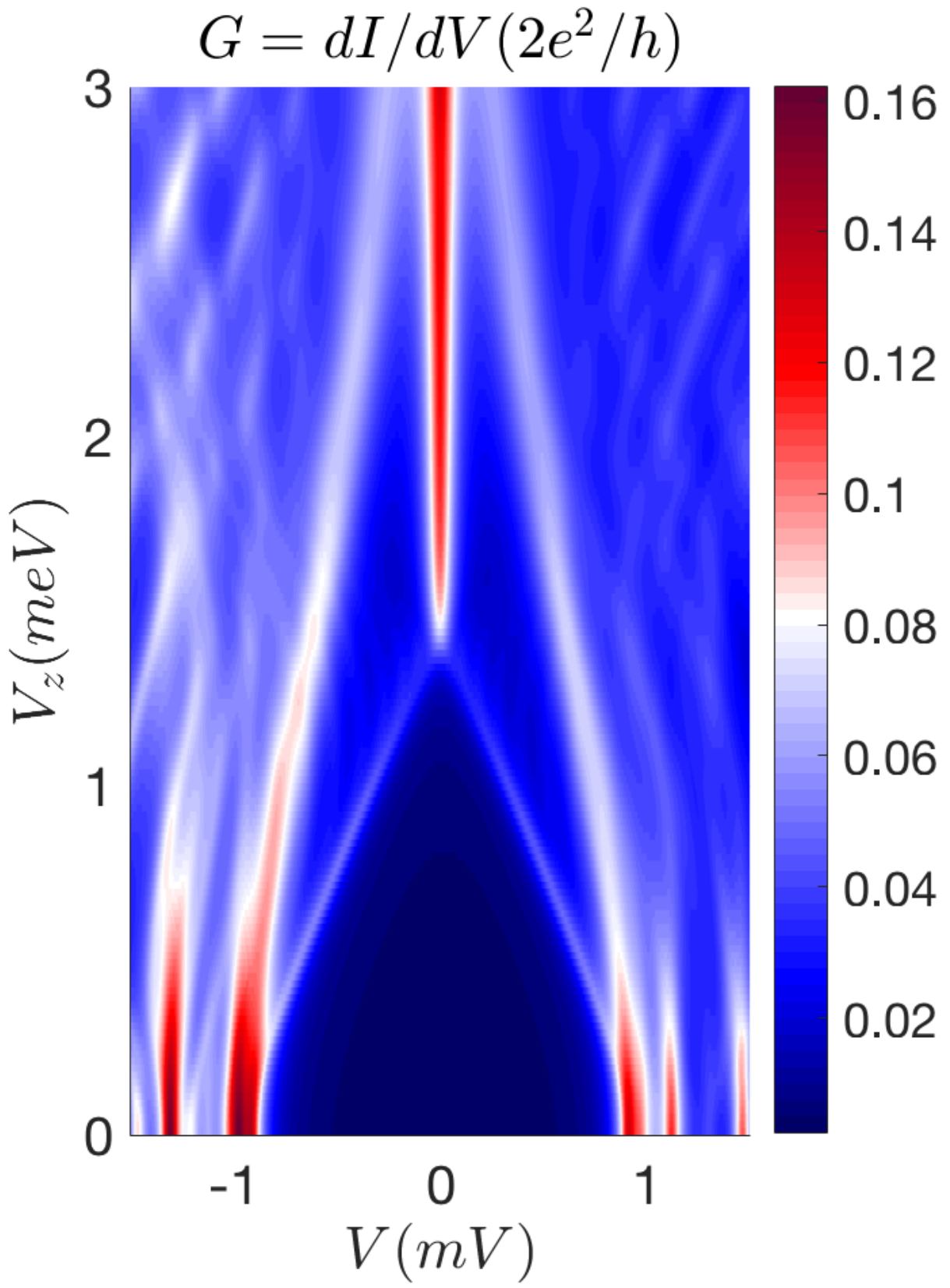}}  
\subfigure[\label{fig:water_zoomTPT}]{
\includegraphics[width=7.5cm, height=11cm]{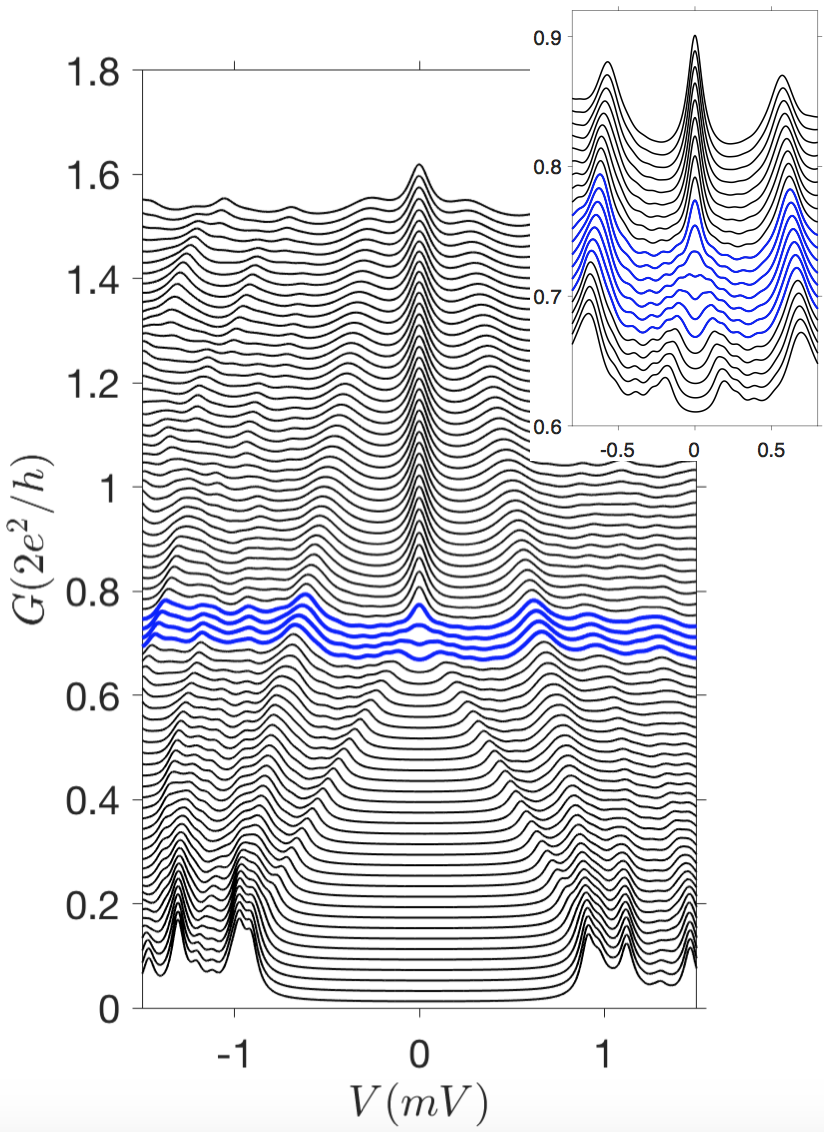}} 
\caption{(color online). (a) Best-fitting conductance. Gate voltage in the lead is assumed to give $E_{\mathrm{lead}}=-$20 meV. The narrow barrier has width $D=20$ nm and height $E_{\mathrm{barrier}}=30$ meV. (b), line cuts from the data in $(a)$ with vertical offsets $0.02\times 2e^2/h$. Inset zooms into the region close to the topological phase transition with vertical offsets $0.01\times 2e^2/h$.}
\label{fig:bestfit}
\end{figure*}

\section{Minimal theory}
We use the following low-energy effective Hamiltonian for the Majorana nanowire~\cite{Sau2010Generic, Lutchyn2010Majorana, Oreg2010Helical}
\begin{align}
H = \left( -\frac{\hbar^2}{2m^*} \partial^2_x -i \alpha_R \partial_x \sigma_y - \mu \right)\tau_z + V_z \sigma_x + \Delta \tau_x - i \Gamma,
\label{Eq:Ham}
\end{align}
where $\sigma_{\mu}(\tau_{\mu})$ are Pauli matrices in spin (particle-hole) space. Some parameters are fixed by experimental measurements~\cite{Zhang2016Ballistic}, e.g. effective mass $m^* = 0.015m_e$, induced superconducting gap $\Delta=0.9$meV, and nanowire length $\sim1.3\mu$m. Zeeman energy is $V_z$[meV] $= 1.2B[T]$, based on an estimation $g_{\mathrm{InSb}} \simeq 40$. The unknown parameters are spin-orbit coupling $\alpha_R$, chemical potential $\mu$, and the phenomenological dissipation parameter $\Gamma$~\cite{Sarma2016How} (which is further discussed below). The lead and barrier are also described by Eq.~(\ref{Eq:Ham}), but without the last two terms on the right-hand side, and with an additional on-site energy $E$ that represents gate voltage. Multi-sub-band effect is introduced by constructing two separate nanowires with different chemical potentials. 

\section{Numerical method}
To calculate the tunneling conductance through the normal-superconductor (NS) junction, we use KWANT~\cite{Kwant}, which is a Python package for numerical calculations on tight-binding models giving the S matrix for scattering regions. We discretize Eq.~(\ref{Eq:Ham}) into a tight-binding model and extract the differential conductance from the corresponding S matrix~\cite{Blonder1982Transition, Setiawan2015Conductance}. 

\begin{figure*}[!htb]
\subfigure[\label{fig:Ta} $T=0.0meV $]{
\includegraphics[width=4cm, height=6.0cm]{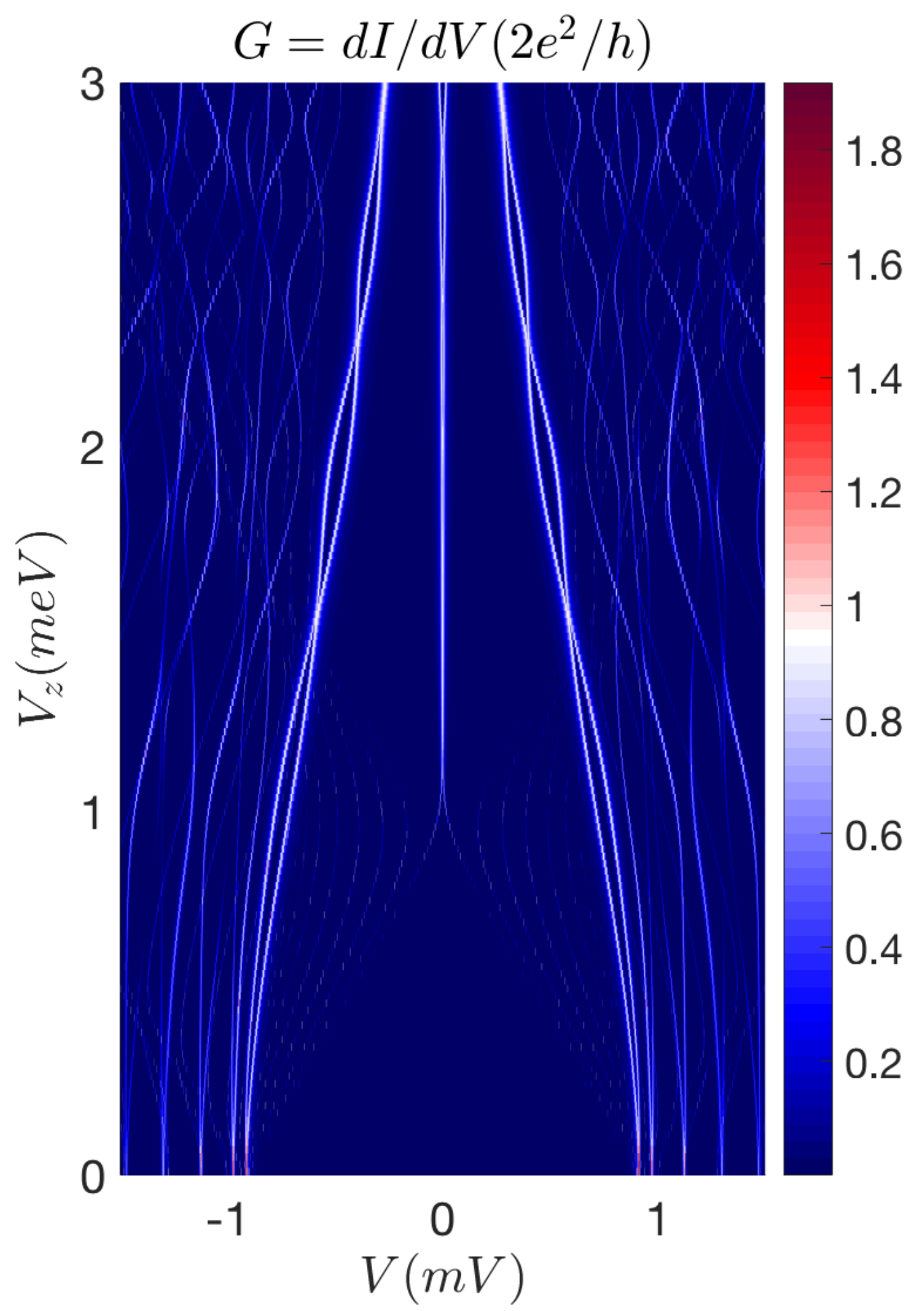}}  
\subfigure[\label{fig:Tb}  $T=0.01meV$]{
\includegraphics[width=4cm, height=6.0cm]{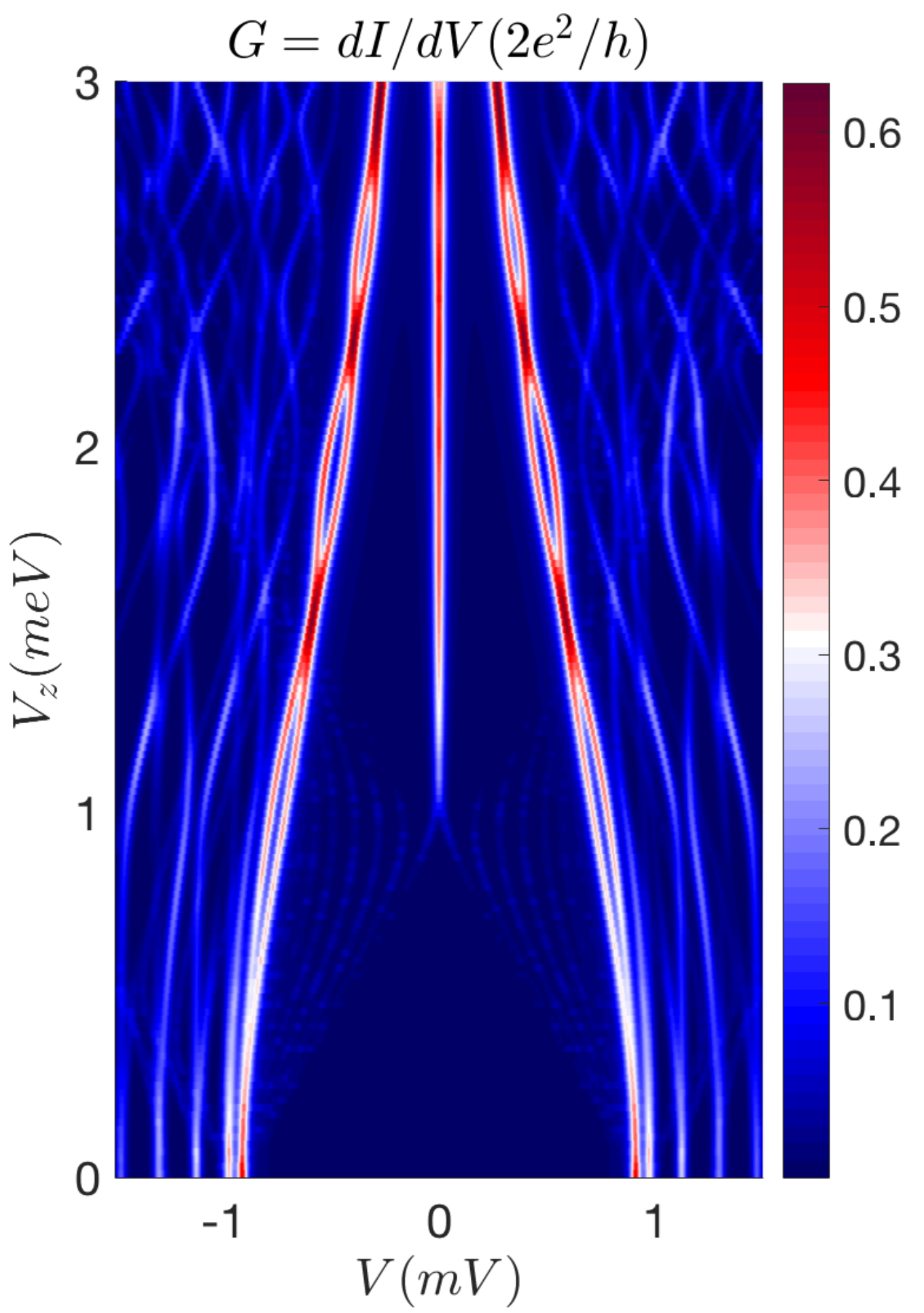}} 
\subfigure[\label{fig:Tc}  $T=0.05meV$]{
\includegraphics[width=4cm, height=6.0cm]{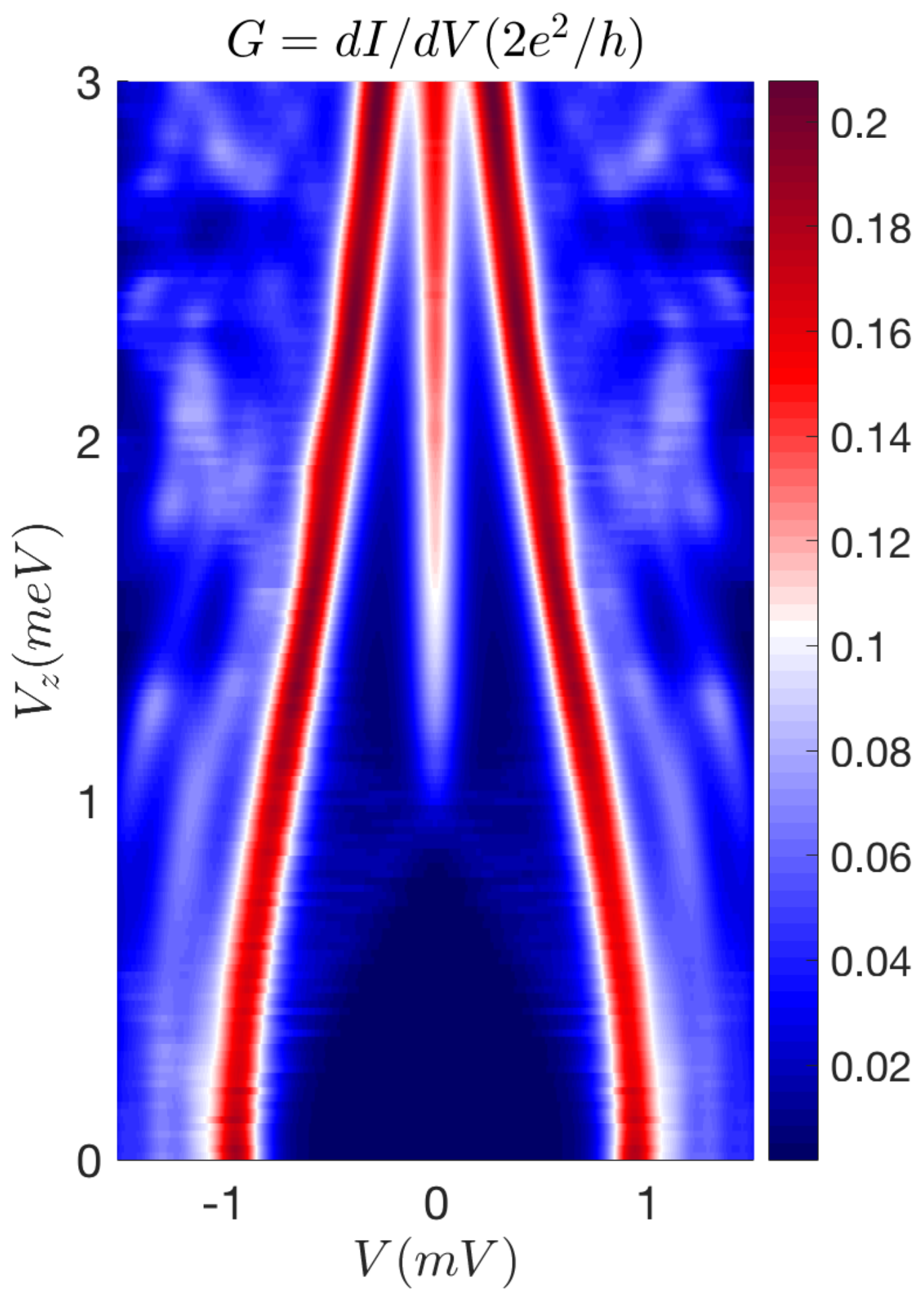}} 
\subfigure[\label{fig:Td}  $T=0.1meV$]{
\includegraphics[width=4cm, height=6.0cm]{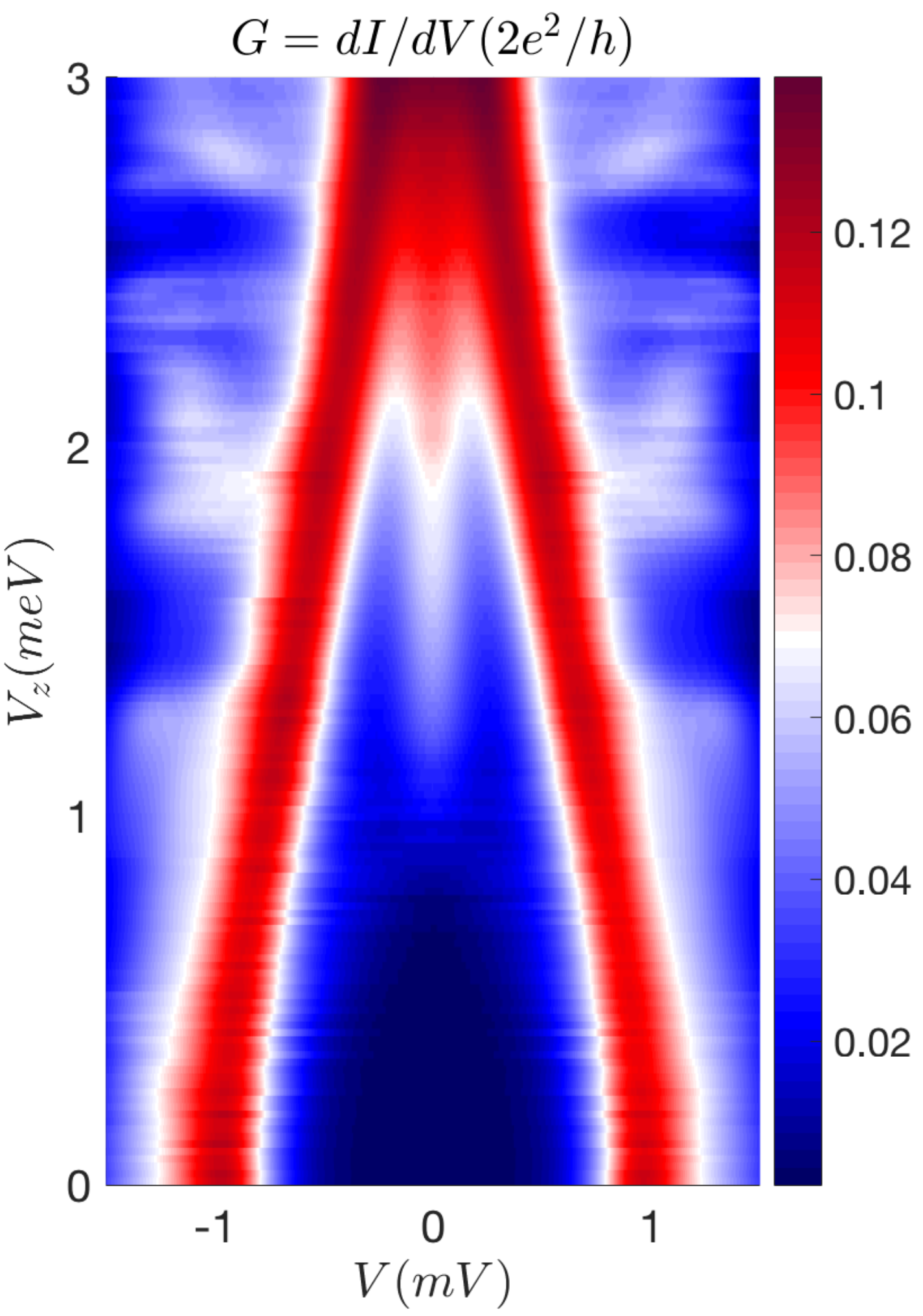}} 
\caption{(color online). Tunneling conductance at different temperatures without dissipation. Finite temperature broadens the ZBCP and lowers its peak value simultaneously, without breaking any p-h symmetry. Chemical potential of the first band is $\mu_1=0$ meV. Only (a) is calculated by KWANT; others are generated by convolution.}
\label{fig:T_effects}
\end{figure*}

\section{Best-fit conductance plot}
Figure 1 shows the calculated conductance for the NS junction with optimal parameters, which agrees well with the data in Ref.~\cite{Zhang2016Ballistic}. The spin-orbit coupling parameter $\alpha_R$, which controls the splitting of the ZBCP in a finite nanowire, is chosen to be as large as $\alpha_R = 0.5$ eV$\AA$, since ``Majorana oscillation'' is not observed in experiments~\cite{Zhang2016Ballistic}. Chemical potentials of the two bands are tuned as $\mu_1 = 1$ meV, $\mu_2 = 5$ meV, such that within the regime $1.3 \lesssim V_z < 3$ meV, only one band is topological. Due to such a difference in Fermi momentum of the two bands, barrier potential affects them differently: increasing barrier width would give larger side peaks, assuming the ZBCP is kept the same. Thus in order to match the data in Ref.~\cite{Zhang2016Ballistic} quantitatively, we choose a narrow barrier. The temperature in Fig.~\ref{fig:bestfit} is chosen to be $T=50$ mK consistent with the quoted temperature in the experiment~\cite{Zhang2016Ballistic}, and changing $T$ to $100$  mK does not change the results in Fig.~\ref{fig:bestfit}(higher-$T$ results are shown in Fig.~\ref{fig:T_effects}).  Dissipation of each band is assumed to depend on Zeeman field: $\Gamma_1 = 0.05(1+0.2 V_z)$ meV, $\Gamma_2 = 0.05(1+ V_z)$ meV such that the side peaks are less obvious at large Zeeman energies, as observed experimentally (other choices for dissipation, including constant $\Gamma$, do not make any qualitative difference). One interesting feature is that a dip in conductance at zero bias grows into a peak when the system undergoes a topological phase transition [blue cutlines in Fig.~\ref{fig:water_zoomTPT}]. This general phenomenon is consistent with experimental observations~\cite{Mourik2012Signatures, Das2012Zero, Deng2012Anomalous, Churchill2013Superconductor, Finck2013Anomalous, Albrecht2016Exponential, Zhang2016Ballistic, Chen2016Experimental}. We also reproduce the finite-field soft gap feature as observed in Ref.~\cite{Zhang2016Ballistic} and other experiments.

\section{Finite temperature}
Finite temperature is one of the mechanisms that can explain the significant discrepancy between the theoretically predicted $T=0$ quantized conductance ($2e^2/h$) and the much lower value observed experimentally. The conductance at finite temperature is computed from the zero-temperature conductance (assuming we neglect the voltage dependence of the barrier) by a convolution with the derivative of Fermi distribution: $G_T(V)=-\int dE G_0(E) f^{'}_T(E-V)$. As shown in Fig.~\ref{fig:T_effects}, with rising temperature, conductance profiles, including the ZBCP, get broadened and peak values go down without breaking any p-h symmetry. In this paper we consider temperature up to $0.1$ meV $\sim$ 1.2 K [e.g., Figs.~\ref{fig:Td} and \ref{fig:GamvsT}] Without dissipation, however, such a ZBCP width is then simply the thermal broadening.

\begin{figure*}[!htb]
\subfigure[\label{fig:Ga} $\Gamma=10^{-4}meV $]{
\includegraphics[width=4cm, height=6.0cm]{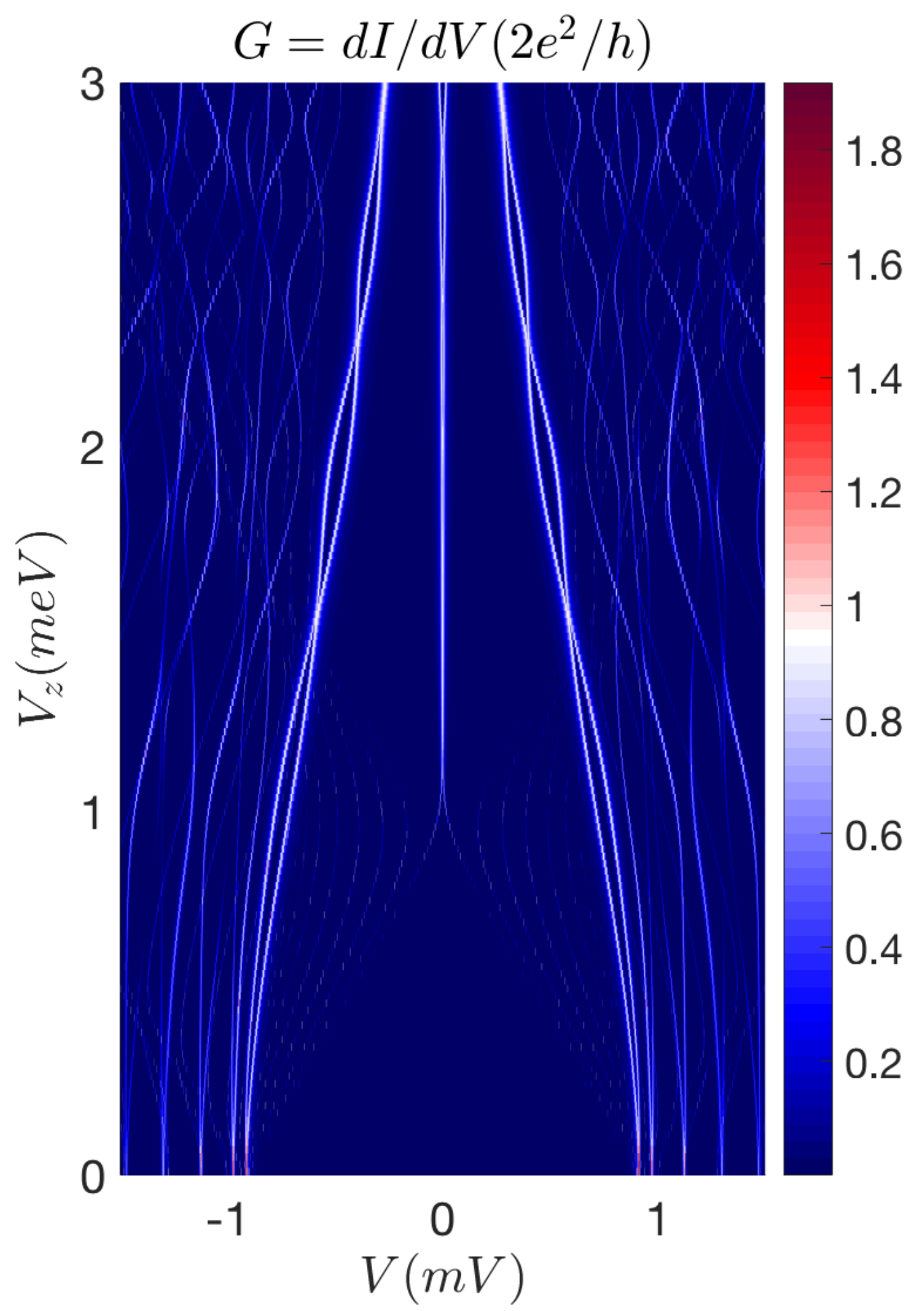}}  
\subfigure[\label{fig:Gb}  $\Gamma=0.01meV$]{
\includegraphics[width=4cm, height=6.0cm]{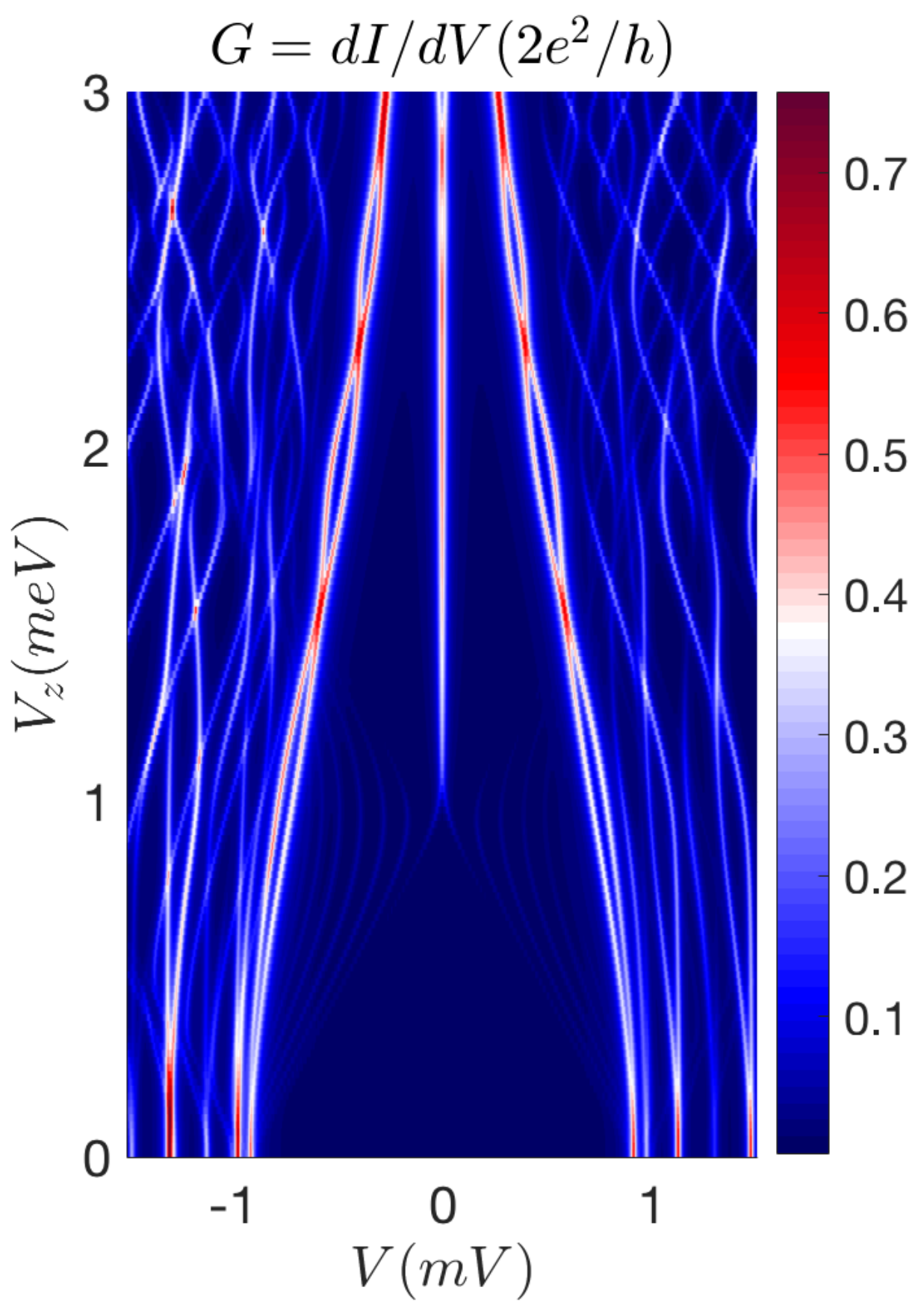}} 
\subfigure[\label{fig:Gc}  $\Gamma=0.05meV$]{
\includegraphics[width=4cm, height=6.0cm]{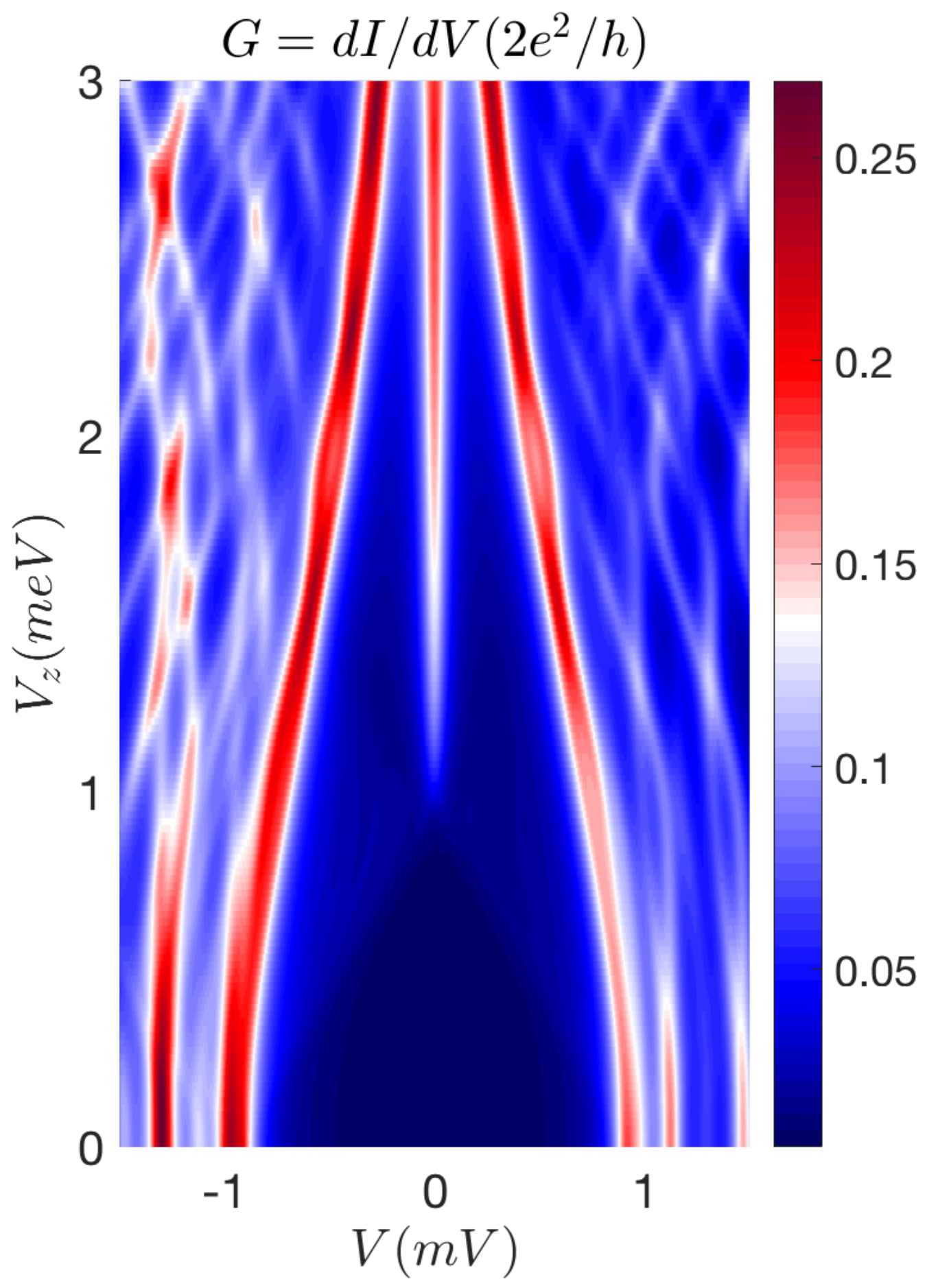}} 
\subfigure[\label{fig:Gd}  $\Gamma=0.1meV$]{
\includegraphics[width=4cm, height=6.0cm]{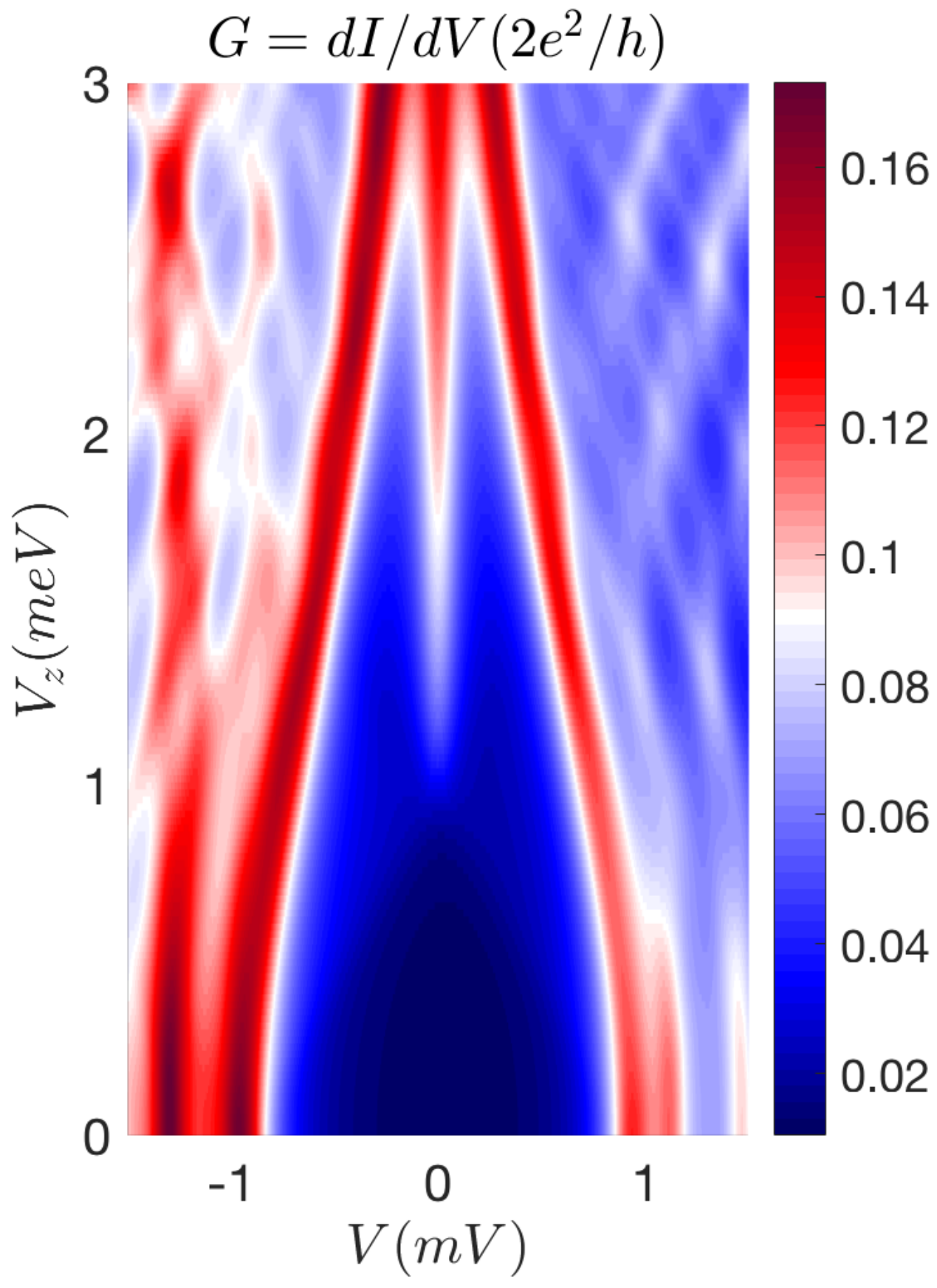}} 
\caption{(color online). Tunneling conductance with various dissipation at zero temperature. Dissipation lowers the peak value of ZBCP and broadens its width. Furthermore, it breaks p-h symmetry in conductance at finite energies. Chemical potential of the first band is $\mu_1=0$ meV. All the four plots are generated by KWANT.}
\label{fig:dissipation_effects}
\end{figure*}

\section{Dissipation}
As we will argue later in the paper, dissipation through a fermionic bath~\cite{Sarma2016How} appears critical to understanding certain features in the conductance data. Physically the dissipation considered here not only includes energy loss but also loss of fermions as in the presence of a fermion bath. At magnetic fields beyond the critical magnetic field, the superconductor becomes populated with vortices containing normal cores that can behave as a fermion bath. Additionally (and more importantly at lower magnetic fields) such dissipation may potentially arise from the combination of disorder and interaction. Disorder can lead to subgap states in the middle of the wire, which would not be visible in conductance. Electrons in the process of Andreev reflections from bound states at the end of the wire can decay into these deeper bound states through the interactions. This effectively leads to dissipation similar to a fermion bath. The parent superconductor itself in the presence of disorder and vortices provides an additional dissipative mechanism. All these microscopic mechanisms are summarized phenomenologically into an imaginary part of the on-site energy (i.e., $\Gamma$) in Eq.~(\ref{Eq:Ham}). Numerical simulations including dissipation are shown in Fig.~\ref{fig:dissipation_effects}: dissipation broadens the conductance profile, including the ZBCP, and lowers their peak values (and also softens the gap somewhat). Furthermore, dissipation introduces p-h asymmetry into the conductance at finite energies, while the ZBCP is still p-h symmetric. This interesting phenomenon can be understood according to Refs.~\cite{Martin2014Noneq, Yazdani1997Probing, Bauriedl1981Electron}, where it is shown that for a tunneling system with a nonequilibrium distribution, the p-h symmetry of the conductance profile is respected only if there is no extra bath (i.e., no dissipation). In contrast, with an extra bath causing dissipation which is much larger than the tunneling amplitude, the result goes back to the standard theory of electron tunneling in the NS junction~\cite{schrieffer1963theory}, i.e., conductance at positive (negative) energy is proportional to electron (hole) density of states at that energy, which is not necessarily p-h symmetric. We believe this is what is going on in the Majorana nanowire experiments where p-h symmetry breaking seems generic. Here we ignore p-h asymmetry caused by the unequal barrier due to voltage bias, since such trivial effect should be minimal for p-h asymmetry at low voltage~\cite{Chen2016Experimental} (and can also be easily experimentally checked). As Fig.~\ref{fig:dissipation_effects} shows, when dissipation is negligible [Fig.~\ref{fig:Ga}], the conductance is p-h symmetric. With increasing dissipation, p-h asymmetry shows up more explicitly until when the dissipation is large enough such that the ratio between conductance at positive and negative biased voltage reaches some limit, which is the ratio of electron and hole weight of the BdG eigenfunction at that energy. However, regardless of dissipation, the ZBCP profile itself is always p-h symmetric, because MZM always has equal electron and hole weights. We therefore conclude that dissipation has qualitatively the same effect on the ZBCP strength (see Fig.~\ref{fig:GamvsT}) as finite temperature: both broaden and lower the ZBCP without breaking its p-h symmetry, and it is thus difficult to disentangle the two effects from the ZBCP. For conductance at finite energies, dissipation produces p-h asymmetry while temperature does not.

\section{Temperature versus dissipation effects on ZBCP}
\begin{figure}
\includegraphics[width=0.45\textwidth]{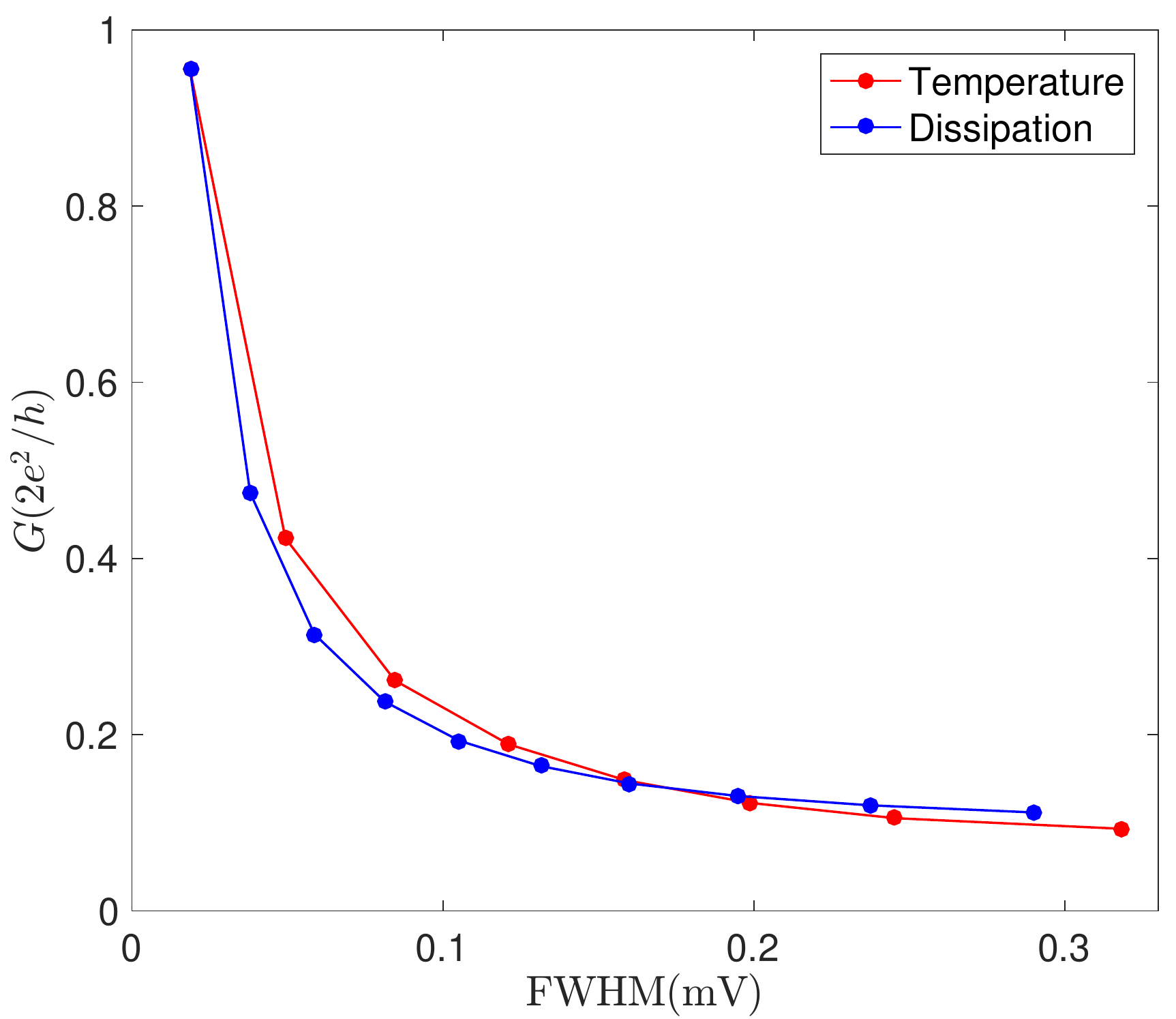}
\caption{(color online). The peak value and FWHM of ZBCP for temperature and dissipation. Data points on the red curve are obtained at increasing temperature but without dissipation, while points on the blue curve are obtained with increasing dissipation at $T=0$. $V_z = 2$ meV, $\mu_1=0$ meV.}
\label{fig:GamvsT}
\end{figure}

Following the previous discussion, Fig.~\ref{fig:GamvsT} gives a quantitative comparison, showing how the peak value and full width at half maximum (FWHM) of ZBCP vary with dissipation and temperature respectively. Both effects give almost identical variation of ZBCP profile, indicating that the huge discrepancy between the quoted temperature ($\sim 50mK$) and the peak value of ZBCP can be explained by dissipation mechanism, since at $T=50 mK$ (without any dissipation) the ZBCP value should be close to $2e^2/h$.

\section{Particle-hole asymmetry at superconducting gap}
While the inclusion of dissipation allows the possibility of p-h symmetry breaking it does not guarantee it, e.g., the conductance in the so-called tunneling (dissipation dominated) limit to a conventional BCS superconductor without spin-orbit or Zeeman fields~\cite{Tinkham1996introduction} is known to be p-h symmetric. However, in the experimental data~\cite{Zhang2016Ballistic}, the SC gap at positive and negative biased voltages shows explicit p-h asymmetry. From the conventional theory of an $s$-wave superconductor, the p-h symmetry at and in the vicinity of the SC gap is due to the pair of Bogoliubov quasi-particles with the same excitation energy above and below the Fermi surface, and the small ratio of $\Delta/\mu$~\cite{schrieffer1963theory}. For the second band with $\mu_2 = 5$meV, the second condition is well satisfied, but its large SC coherence length would bring in significant finite-size effect. Therefore the quasi-particle pairs might not have the same excitation energy, causing p-h asymmetry at the order of $(\xi/L)$. Put in another way, the p-h asymmetry at the SC gap arises because dissipation is less than the level spacing of the finite nanowire. In addition, the way of p-h symmetry breaking is random, i.e., either the electron or hole part could have larger contribution, depending on the relative position of the pair of quasi-particle excitations. Based on these arguments, the p-h asymmetry of the SC gap in the second band should decrease with increasing nanowire length. For the first band ($\mu \sim 1$ meV), its large SC gap compared to the Fermi energy and the missing of the excitation branch below the Fermi surface at some threshold energy both can cause p-h asymmetry.

\section{Conclusion}
Through realistic simulations of Majorana nanowires and detailed comparison with recent experiments~\cite{Zhang2016Ballistic} we have identified dissipation, temperature, multi-sub-band, and finite barrier as the important physical mechanisms controlling MZM tunneling conductance properties. Our theoretical results agree well with recent experimental data including the puzzling observation of the breaking of the particle-hole symmetry.

\begin{acknowledgements}
We thank F. Hassler, M. Wimmer, and A. Akhmerov for insightful comments. This work is supported by Microsoft and LPS-MPO-CMTC.
\end{acknowledgements}

\bibliography{BibMajorana}

\end{document}